\documentclass[aps, prx, superscriptaddress]{revtex4-2}

\usepackage{amssymb,amsmath,amsfonts,latexsym, mathrsfs}
\usepackage{graphicx}
\usepackage{stmaryrd}
\usepackage{soul}

\newcommand{\genL}{\mathcal I_\text{gen}}

\begin{document} 
\title{Physics-informed Machine Learning Analysis 
for Nanoscale Grain Mapping by Synchrotron Laue Microdiffraction}
 
\author{Ka Hung Chan}%
\affiliation{Department of Mechanical and Aerospace Engineering, The Hong Kong University of Science and Technology, Hong Kong}
\affiliation{Advanced Light Source, Lawrence Berkeley National Laboratory, Berkeley, United States}

\author{Xinyue Huang}%
\affiliation{Department of Mechanical and Aerospace Engineering, The Hong Kong University of Science and Technology, Hong Kong}

\author{Nobumichi Tamura}%
\affiliation{Advanced Light Source, Lawrence Berkeley National Laboratory, Berkeley, United States}

\author{Xian Chen}
\email{xianchen@ust.hk}
\affiliation{Department of Mechanical and Aerospace Engineering, The Hong Kong University of Science and Technology, Hong Kong}

\date{\today}


\begin{abstract}
\noindent
Understanding the grain morphology, orientation distribution, and crystal structure of nanocrystals is essential for optimizing the mechanical and physical properties of functional materials. Synchrotron X-ray Laue microdiffraction is a powerful technique for characterizing crystal structures and orientation mapping using focused X-rays. However, when grain sizes are smaller than the beam size, mixed peaks in the Laue pattern from neighboring grains limit the resolution of grain morphology mapping. We propose a physics-informed machine learning (PIML) approach that combines a CNN feature extractor with a physics-informed filtering algorithm to overcome the spatial resolution limits of X-rays, achieving nanoscale resolution for grain mapping. Our PIML method successfully resolves the grain size, orientation distribution, and morphology of Au nanocrystals through synchrotron microdiffraction scans, showing good agreement with electron backscatter diffraction results. This PIML-assisted synchrotron microdiffraction analysis can be generalized to other diffraction-based probes, enabling the characterization of nanosized structures with micron-sized probes.
\end{abstract}

\maketitle

\section{Introduction}

Morphological configurations of polycrystals, including phase interfaces, grain boundaries, orientation distribution, and textures, play a profound role in the mechanical and physical properties of functional materials. The nanoscale grains and precipitates underlie disruptive technological advancements such as solar cells \cite{insitu-perovskite-2023, insitu-perovskite-2024}, plasmonic devices \cite{plasmonic2015, plamonic2021}, and thermoelectric devices\cite{thermoelectric2018, thermoelectric2021}. A comprehensive analysis of the morphological features of grains and heterogeneous phases is necessary for the development of novel materials with tailored properties. Laue microdiffraction experiments are widely utilized to characterize crystallographic information, grain orientations, localized strains, heterogeneous phases, and precipitates, but their spatial resolution is fundamentally limited by the size of the focused X-ray probe \cite{ALS2016}.

Nanocrystalline materials with grain sizes below 500 nm are typically difficult to detect using a micron-sized polychromatic X-ray beam, even when generated by a synchrotron source. Recent advances have pushed the optical focusing limit of photon flux to achieve submicron beam sizes (i.e., 300 nm) \cite{APS2011, ESRF2011}, but resolving grains at the nanoscale remains challenging due to the Laue pattern consisting of peaks diffracted from multiple neighboring grains. While new and better X-ray optics might be able to solve this issue, it is not ideal for existing synchrotron facilities. Firstly, ultra-precise optical mirror that produce a 100 nm beam are costly and require substantial modifications to the beamline. \cite{beam_stability2008, nanoprobe_2017, microprobes_2018}. These mirrors are also more difficult to maintain and align, as their performance typically degrades over time. Not all facilities can afford these upgrades. Additionally, even with a very small beam, the problem of X-ray penetration remains. 

Conventional X-ray crystallographic analysis relies on Laue indexing for a set of diffracted peaks from a single orientation \cite{tamura2014xmas}. However, this method is not sufficient to resolve grains with sizes smaller than the beam size, as one Laue pattern could contain many orientations. Even advanced algorithms like XMAS developed by the Advanced Light Source (ALS) \cite{tamura2014xmas}, and LaueTools and LaueNN developed by the European Synchrotron Radiation Facility (ESRF) \cite{lauetools, laueNN}, which can distinguish the superposition of Laue indexing for multiple grains, have their limitations. For instance, LaueNN, a sophisticated neural network-based algorithm, can refine the indexing for about 10 grains, but this is still not enough to resolve nanosized grains. Moreover, such sequential indexing procedures are computationally intensive and time-consuming.

To address the challenges of synchrotron Laue microdiffraction experiments and their analysis approaches for nanocrystals or nanocrystalline functional materials, a fundamental question arises: Is it necessary to obtain Laue indexing for each pattern if the ultimate goal is to understand the material's morphological configurations? Laue patterns diffracted by a synchrotron white beam with a sufficiently wide energy bandpass contain hidden features that can be directly utilized for spatial segmentation of grains with different orientations. Based on our previous work on indexing-free analysis of synchrotron Laue diffraction \cite{song2019}, the Laue pattern can be encoded into 256 latent features by the convolutional neural network (CNN) autoencoder CANON \cite{song2019}. Unlike conventional methods, CANON skips the Laue indexing procedure and directly generates the grain map from the learned latent features. By reducing each image to a feature vector in a lower-dimensional latent space, this method significantly accelerates data processing, especially when handling large datasets. While the CANON analysis package shows promise in extracting hidden features from Laue patterns and bypassing the indexing process, its capability to resolve grains smaller than the X-ray beam size remains uncertain. The potential of CANON to analyze Laue patterns with superpositions of many nanoscale grains without indexing and to identify the primary grain among numerous satellite grains exposed by micron-sized X-ray beams is intriguing. Yet, the superposition of multiple diffractions results in complex, nonlinear combinations of features that the autoencoder struggles to disentangle. To address this challenge, incorporating additional physical information, such as peak intensity distribution, could enhance the machine learning algorithm's ability to resolve features within Laue patterns. This approach may further increase the spatial resolution of micron-sized X-ray beams, enabling the determination of submicron to nanoscale grain morphological distributions. 

In this paper, we present a physics-informed data filtering and pooling algorithm designed for feature engineering to effectively isolate diffracted peaks from primary orientations in Laue patterns. The proposed algorithm generates a set of refined patterns that concentrate the peak intensities corresponding to the primary nano-grain within the exposure domain, which is illuminated by a focused white beam synchrotron X-ray. This method systematically removes stray peaks originating from less-exposed grains, thereby facilitating the feature learning process by CANON. The processed Laue patterns markedly enhance the orientation resolution of local nano-grains, a critical factor for subsequent analysis using the Physics-Informed Machine Learning (PIML) framework. We validate our PIML approach through its application to a nanocrystalline gold thin film deposited on a single-crystal Si substrate. The thin film exhibits a thickness of 1~$\mu$m and features grain sizes on the nanometer scale.

\section{Experiment and method}
\subsection{Setup and philosophy of X-ray microdiffraction for grain mapping}
\begin{figure}
    \centering
    \includegraphics[width=0.95\textwidth]{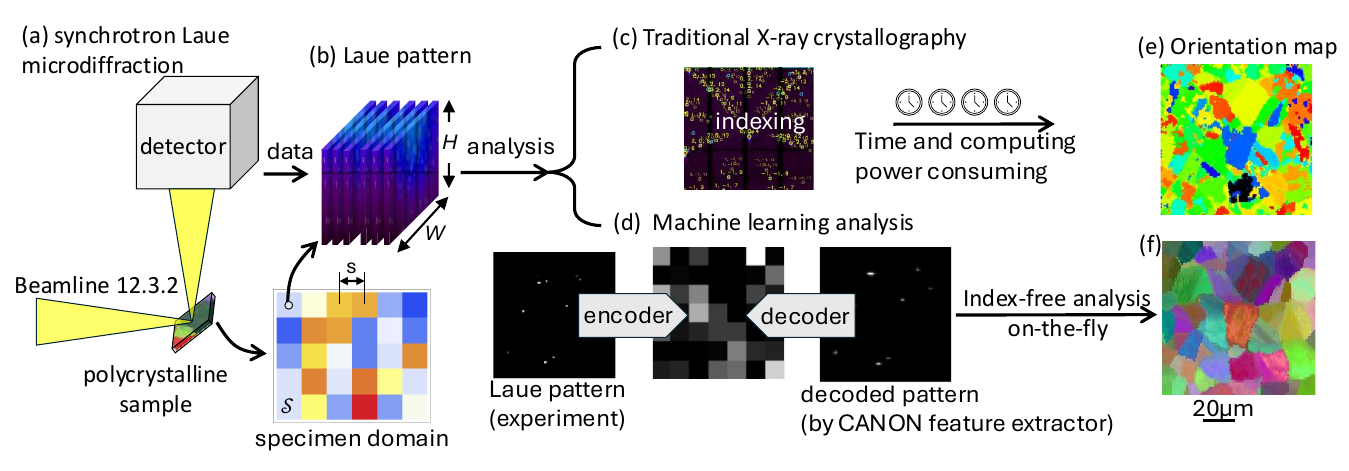}
    \caption{\textbf{Schematics of the synchrotron X-ray Laue microdiffraction experiment} (a) Experiment in beamline, which generates a sequence of (b) Laue patterns corresponding to the 2D specimen domain $\mathcal S$ with a scanning step distance $s$. (c) to (e) illustrate the traditional data analysis procedure by X-ray crystallography, while (d) to (f) elaborates on the machine learning approach to analyze the Laue microdiffraction data through feature extraction and unsupervised learning.}
    \label{fig:expr}
\end{figure}

The standard equipment setup and data analysis of synchrotron X-ray microdiffraction experiment is illustrated in Fig.\ref{fig:expr}. We perform a 2D raster scan on the specimen domain defined as
\begin{equation}\label{eq:sdomain}
\mathcal S(n_i, n_j) = \{(x, y): (x, y)=(x_0,y_0)+s(n_i,n_j) \text{ for } (n_i, n_j)\in \mathbb Z^2\}
\end{equation}
where the vector $(x_0, y_0)$ specifies the starting location on the sample surface associated with the left-upper corner in the specimen domain, the value $s > 0$ denotes the scanning step size by the focused X-ray beam, and the range of integer tuples $(n_i, n_j)$ underlies the total number of Laue patterns collected in each of the microdiffraction experiments. Mathematically, the microdiffraction experiment is a mapping $\mathcal{I}: \mathcal S \to \mathbb R^{H\times W}$ where $H$ and $W$ are the pixelate dimensions of a Laue pattern image, in Fig.\ref{fig:expr}(b). For  $(n_i, n_j) \in [0, N_x - 1] \times [0, N_y - 1]$, the specimen domain is mapped to $N_x N_y$ Laue pattern images. For $(x, y) \in \mathcal S$, the Laue pattern can be expressed as $\mathcal I(x, y) \in \mathbb R^{H\times W}$. 

If the grain size is larger than the X-ray beam size, the scanning step $s$ can be set to be at least the beam size, then Laue patterns in the scanning sequence are sufficient to distinguish the orientation difference across the grain boundaries by traditional X-ray crystallographic analysis, in Fig.\ref{fig:expr} (b) $\to$ (c), finally generates an orientation map for polycrystal domain in Fig.\ref{fig:expr} (e). Note that the grain boundaries resolved by Laue indexing are not spatially precise as the Laue patterns collected near the grain boundary consist the $(hkl)$ peaks diffracted by both neighboring grains. By machine learning (ML) assisted analysis \cite{song2019}, as seen in Fig.\ref{fig:expr} (b) $\to$ (d), the grain boundaries in the scanned domain can be resolved more precisely, as seen in Fig.\ref{fig:expr} (f). 

If the grain size is smaller than the X-ray beam size but larger than the scanning step size, the diffraction peaks in neighboring Laue patterns will overlap. For example, at the microdiffraction beamline 12.3.2 at Advanced Light Source, Lawrence Berkeley National Lab, the precise motorized stage can achieve a scanning step of $s \geq 100$ nm with the X-ray focused to $1 \mu$m. If the grain size is approximately 500 nm, the neighboring Laue patterns along the scanning sequence may consist of peaks diffracted by the same grain, illustrated in Fig.\ref{fig:datapooling} (a). In this case, direct crystallographic analysis may not resolve the spatial boundaries between grains with different orientations. Even if we can index the highly entangled Laue pattern with advanced indexing algorithms \cite{tamura2014xmas, laueNN, patternmatching2022, indexing_optimal_transport2024}, the intensity profiles shown in Fig.\ref{fig:datapooling} (b) reveal that the Grain 1 (red) and Grain 2 (blue) are still unsolvable. Nevertheless, the geometric features and intensity profiles of each Laue peak possess subtle differences. Such subtle differences can be used to differentiate the spatial domain with proper \emph{learning} of the latent features for the Laue pattern.

\subsection{Physics-informed peak filtering and pooling method}

\begin{figure}[h]
    \centering
    \includegraphics[width=0.4\linewidth]{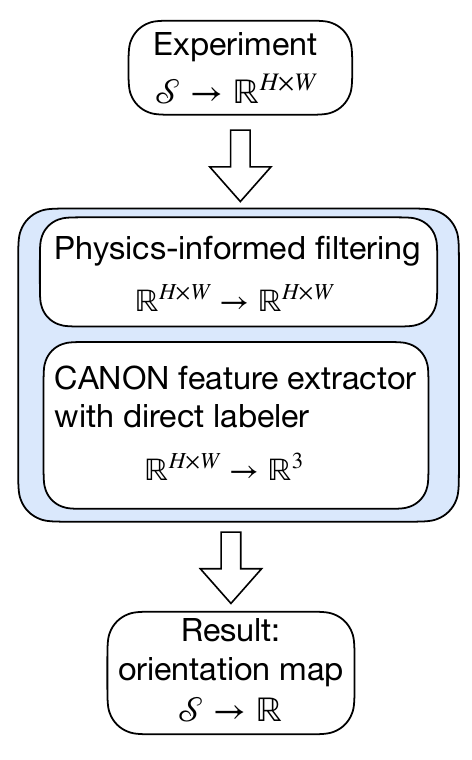}
    \caption{\textbf{Physics-informed machine learning workflow for grain mapping by synchrotron X-ray Laue microdiffraction.}}
    \label{fig:wf}
\end{figure}

Utilizing the intensity amplitude and profile of each Laue peak instead of indexing the entire Laue pattern, the illuminated portion of each grain exposed by the micron-sized X-ray beam can be revealed by a physics-informed machine learning workflow, illustrated in Fig.\ref{fig:wf}. For every spatial location $(x, y) \in \mathcal S$, the experimental Laue pattern $\mathcal I(x, y)$ can be mapped to an artificial Laue pattern
\begin{equation}\label{eq:gen}
    \genL: \mathcal I(\mathcal S) \to \mathbb R^{H\times W}.
\end{equation}

The algorithm to calculate $\genL$ involves two main steps. To enhance the spatial resolution of nanoscale grain mapping, we implemented an adaptive mask computation strategy for filtering Laue diffraction peaks. In the neighborhood of each Laue pattern $\mathcal I(x, y)$ within a radius of $(s, s)$ for a Laue microdiffraction scan, we identified the peak with the highest intensity, $i_{max}$. For all peaks in $\mathcal I(x, y)$, we block the peaks if their intensities are less than 80\% of $i_{max}$. This approach effectively suppresses weaker, less significant peaks originating from satellite grains while preserving the dominant reflections from the primary grain in the exposure domain. This computed mask helps filter out minor peaks that only consist of the primary reflections with the highest intensities within a neighborhood of the measured Laue pattern.

\begin{figure}[h]
    \centering
    \includegraphics[width=0.8\textwidth]{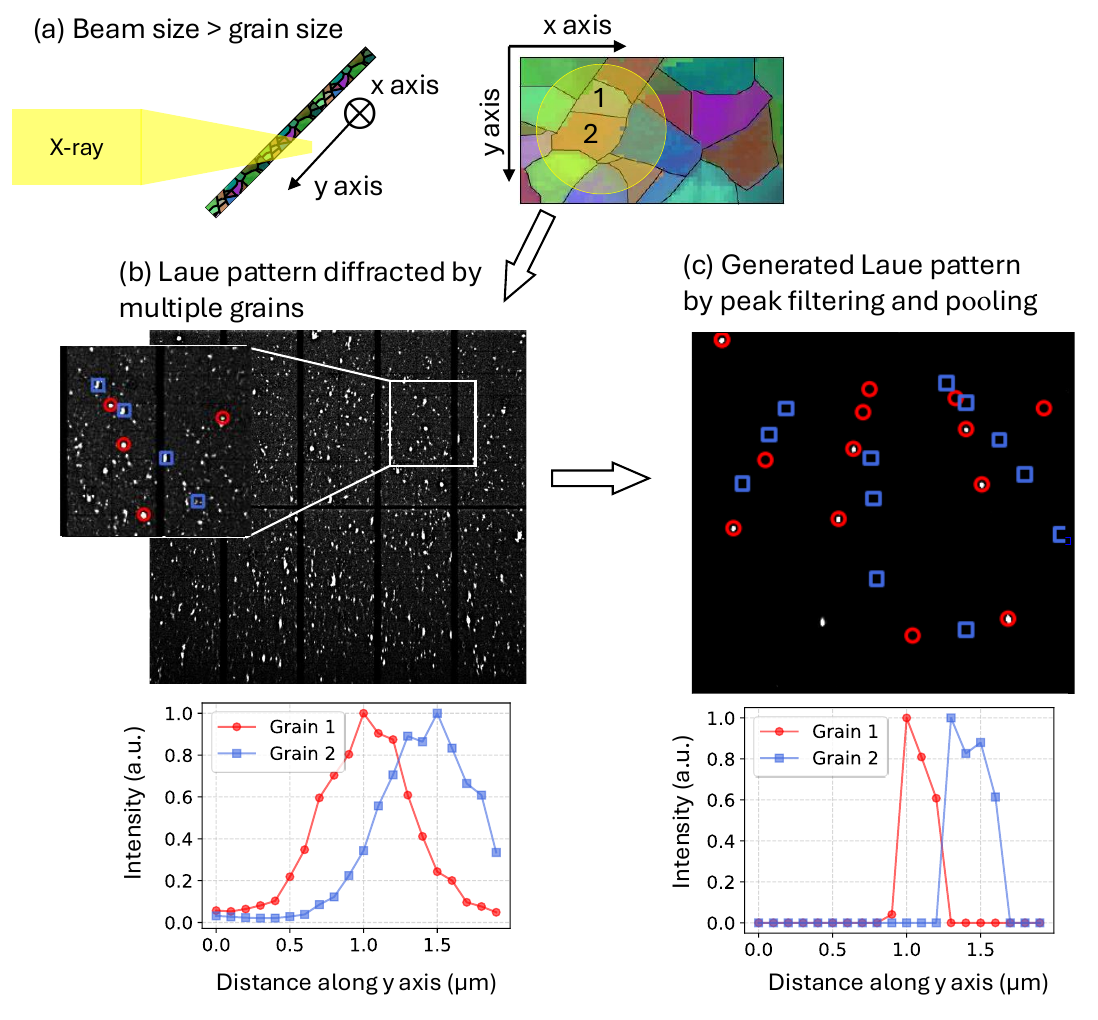}
    \caption{\textbf{Comparison between the Laue pattern collected from experiment and the generated Laue pattern by peak pooling algorithm.} (a) Illustrates the incident X-rays with beam size greater than the grain size. (b) Original Laue pattern diffracted by multiple grains with different orientations. The intensity profiles of grain 1 and grain 2 are spatially unsolvable. (c) Generated artificial Laue pattern by \eqref{eq:gen} mainly shows the reflections from grain 1. The intensity profile of grain 1 and grain 2 are better spatially solvable.}
    \label{fig:datapooling}
\end{figure}

Following the mask computation, we proceed with peak filtering and pooling, an algorithm that aligns with the nature of X-ray diffraction by crystals. By leveraging the inherent characteristics of diffracted peaks, this method effectively distinguishes the diffraction patterns from neighboring grains. As a result, the complex Laue patterns are significantly clarified, making them much cleaner and more suitable for further analysis. Fig.\ref{fig:datapooling} (c) demonstrates the generated artificial Laue pattern after peak filtering and pooling, highlighting that Grain 1 (red) predominantly contributes to the primary reflections, in contrast to Grain 2 (blue). The intensity profiles of the generated artificial pattern $\mathcal I_\text{gen}$ reveal improved resolution of the two grains within the spatial specimen domain along the y-axis.

We used the Laue pattern autoencoder, CANON \cite{song2019}, to extract latent features from the generated Laue patterns. Subsequently, we applied an indexing-free machine learning algorithm \cite{song2019} to segment the specimen domain $\mathcal{S}$ based on the pairwise distances in the reduced-dimensional latent feature space. Particularly, the feature vectors are reduced to 3 dimensions ($\mathbb R^3$) by Principle Component Analysis (PCA). We used each component of the feature vector to represent one of the red/blue/green colors, by which we encode the grain map from the topological clustering results, as elaborated in Fig.\ref{fig:wf}.

\section{Results and discussion}

We demonstrated our Physics-Informed Machine Learning (PIML) method to characterize the nanocrystalline Au thin film on a silicon substrate. The film was fabricated by electron beam evaporation on a cleaned single-crystal silicon (Si) substrate, resulting in a thin film with a thickness of approximately 1 $\mu$m and grain sizes in the nanometer range. We conducted synchrotron X-ray Laue microdiffraction on the sample surface at Beamline 12.3.2 of the Advanced Light Source (ALS), Lawrence Berkeley National Laboratory. The focused X-ray beam had a nominal beam size of 1 $\mu$m. A two-dimensional scan using a polychromatic X-ray beam in the energy range of 6 keV to 22 keV was performed with a step size of 100 nm, systematically capturing Laue diffraction patterns across the specimen domain.

Since the thin layer of Au does not fully diffract all X-rays, the Laue patterns from the specimen domain exhibited very strong silicon reflections. To address this, we conducted a background filtering process to suppress the silicon reflections, ensuring clearer patterns for analysis. This preprocessing step was essential before applying the PIML analysis for nanocrystalline mapping.

\subsection{Grain mapping of Au nanocrystalline thin film}

\begin{figure}[h]
    \centering
    \includegraphics[width=0.9\textwidth]{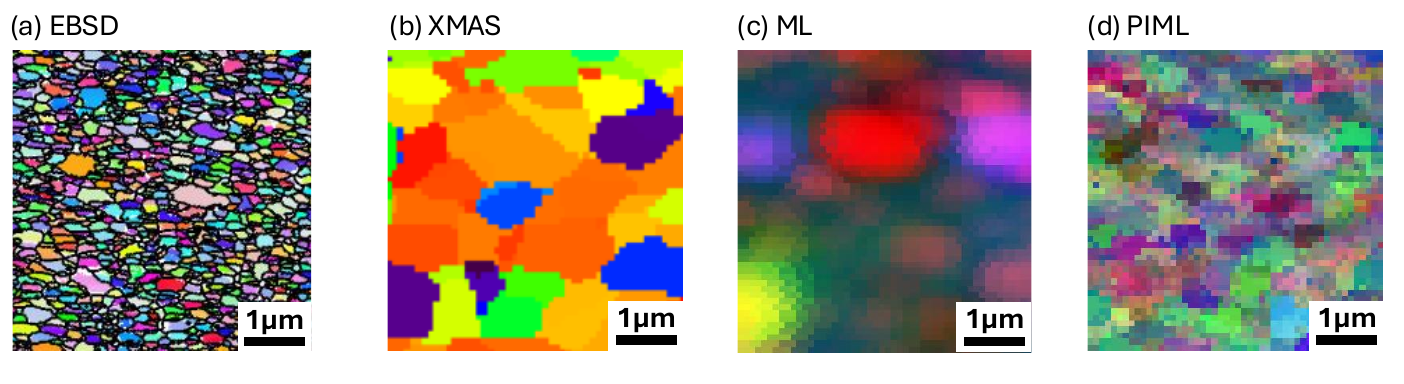}
    \caption{\textbf{Grain map generated by different methods.} (a) Orientation map by EBSD (b) Orientation map by X-ray Laue microdiffraction analyzed by XMAS software. Grain topographic map from the same microdiffraction dataset by (c) indexing-free machine learning algorithm and (d) PIML algorithm.}
    \label{fig:grain_map}
\end{figure}

Electron BackScatter Diffraction (EBSD) was performed on the same specimen using a JEOL JSM-7800F Scanning Electron Microscope (SEM). The scan, conducted with a 25 nm step size over a 20 µm × 20 µm area, utilized a 20 kV high-energy electron beam to differentiate grains via backscattering. The interaction of the electron beam with the crystal lattice generated characteristic Kikuchi patterns, which were analyzed using the point group $m3m$ to create the orientation map shown in Fig. \ref{fig:grain_map}(a). Post-processing with AZtecCrystal software provided precise grain orientations and morphological details, including grain size and orientation distributions, represented by different colors. While the EBSD-characterized region does not perfectly overlap with the area measured by X-ray microdiffraction, the grain morphologies from both experiments are comparable due to the random grain distribution in the Au thin film.

We utilized the synchrotron X-ray Microdiffraction Analysis Software (XMAS) \cite{tamura2014xmas} to index each of the Laue patterns using traditional crystallographic analysis, resulting in the grain mapping shown in Fig.\ref{fig:grain_map}(b). However, the Laue indexing process was insufficient in accurately determining grain boundaries among nanosized grains. This limitation arises from the fact that each Laue pattern consists of diffraction peaks from many satellite grains, and the micron-sized X-ray beam cannot spatially resolve the grain boundaries. Consequently, XMAS struggled to distinguish nanosized grains, leading to incomplete or blurred grain boundary definitions.

For comparison, Fig.\ref{fig:grain_map}(c) shows the grain mapping produced by an indexing-free machine learning (ML) algorithm without physics-informed filtering and pooling processes. While the ML approach offered an alternative method, it similarly failed to identify nanosized grains, highlighting the complexity of accurate grain boundary detection at this scale.

In contrast, the grain mapping generated by the physics-informed machine learning (PIML) method, shown in Fig.\ref{fig:grain_map}(d), successfully reveals fine grains at scales of several hundred nanometers. This result aligns closely with the morphology obtained from EBSD in Fig.\ref{fig:grain_map}(a), demonstrating the effectiveness of the PIML approach in resolving nanoscale features. Additionally, the nanocrystalline Au grains exhibit noticeable anisotropy, with most grains elongated along the horizontal direction. This anisotropic feature is evident in both Figs. \ref{fig:grain_map}(a) (EBSD) and \ref{fig:grain_map}(d) (PIML), further confirming the consistency between these methods.

\subsection{Grain morphology analysis}

\begin{figure}[h]
    \centering
    \includegraphics[width=\textwidth=0.7]{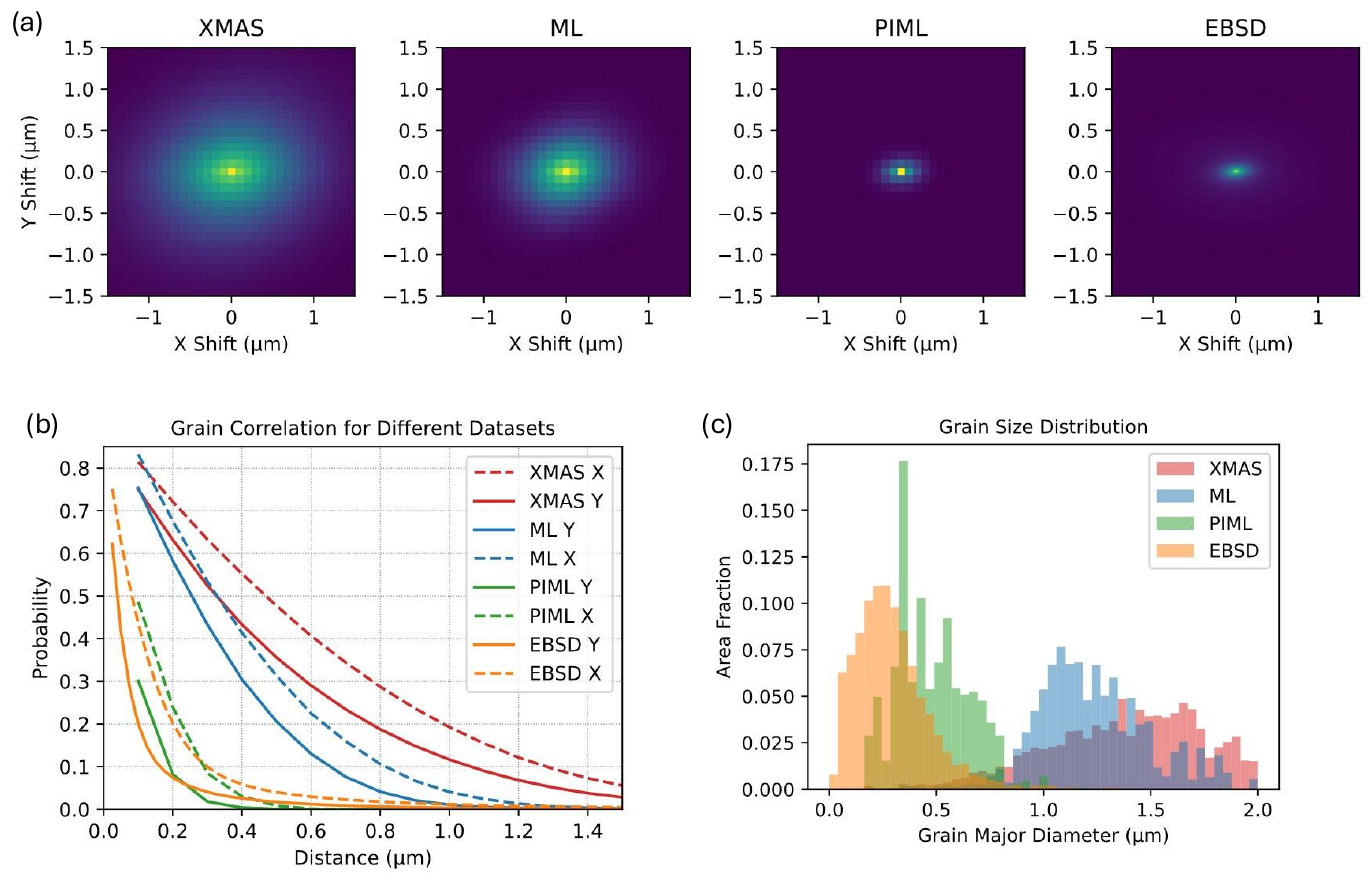}
    \caption{\textbf{Quantitative assessments of grain segmentation by different analysis methods.} (a) two-point correlation of the grain maps given by different methods. (b) Grain size distribution along X and Y directions corresponding to (d) expectations and variances of grain morphologies. }
    \label{fig:grain_correlation}
\end{figure}

A quantitative comparison among the grain maps given by different methods in Fig.\ref{fig:grain_map} was performed using a two-point correlation analysis, as illustrated in Fig.\ref{fig:grain_correlation}. The two-point correlation function is a statistical measure that calculates the probability of any two points within the specimen domain belonging to the same grain. To assess grain size using this function, we calculate the probability for each pair of points and plot it as a function of the distance between the points. The rate at which this function decays with distance provides information about the grain size: a diffusive decay indicates larger grains, while a sharp decay suggests smaller grains. By fitting the correlation function, we can extract parameters that describe the anisotropy of grain morphology (Fig.\ref{fig:grain_correlation}a) and grain size distribution including the mean grain size (Fig.\ref{fig:grain_correlation}b) and variance (Fig.\ref{fig:grain_correlation}c).

The grain morphology obtained by XMAS and ML without physics-informed filtering exhibit a Gaussian shape with a Full Width at Half Maximum (FWHM) of approximately ${1\mu\text{m}}$, close to the size of the X-ray beam, indicating that the spatial resolution of these methods is limited by the beam size. In contrast, the grain morphology given by physics-informed ML is similar to that given by the EBSD scan, with a FWHM of approximately ${0.3 \mu\text{m}}$. It suggests a better spatial resolution beyond the beam size limit. 

Both the grain maps characterized by EBSD and analyzed by PIML algorithms exhibit grain anisotropy, with higher correlation observed in the x-direction compared to the y-direction. This indicates that the grains are predominantly elongated horizontally (i.e., along the x-direction). The PIML method effectively captures this grain anisotropy compared to other analysis methods. The grain size distribution, expectation, and variance of the grain map given by PIML match the EBSD data very well. Specifically, both methods show a Poisson distribution, with a mean grain size of approximately 300 nm for EBSD and 480 nm for the PIML method.

While the PIML algorithm improves the effective spatial resolution, it still slightly overestimates the mean grain size compared to that measured by EBSD. One possible reason for this discrepancy is the residual influence of the beam size and step size. Additionally, the Laue mask used in \eqref{eq:gen} for physics-informed filtering plays a crucial role in determining the grain boundaries. Furthermore, since Laue scanning takes several hours to complete, experimental noise and variations in peak intensity measurements could also contribute to this discrepancy.

\section{Conclusion and outlook}     

In conclusion, the physics-informed machine learning method enhances the spatial resolution of Laue microdiffraction characterization for nanocrystalline polycrystals. While powder diffraction with a large monochromatic beam can provide accurate average grain size and texture information, Laue microdiffraction offers additional insights such as grain boundary morphology, orientation distribution and subgrain information. Moreover, Laue microdiffraction has advantages over EBSD, including strain sensitivity, the absence of the need for additional sample preparation and vacuum compatibility. This improvement significantly impacts materials characterization, as accurate knowledge of grain size distributions and orientations is crucial for understanding material properties. Future work could focus on optimizing the physics-informed hyperparameters and exploring the algorithm's applicability to various materials under different experimental conditions and crystal symmetries.

\begin{acknowledgements}
X. C. and K. H. C. thank the financial support under GRF Grants No. 16203021, 16204022 and No. 16203023 by Research Grants Council, Hong Kong. This research used resources of the Advanced Light Source, which is a DOE Office of Science User Facility under contract no. DE-AC02-05CH11231. K. H. C. was supported in part by an ALS Doctoral Fellowship in Residence.
\end{acknowledgements}

%

\bibliographystyle{apsrev4-2}
\bibliography{iucr} 

\end{document}